\documentclass[aps,prb,preprint,superscriptaddress]{revtex4-1}

\usepackage{amsmath, amssymb}

\newcommand{\commentoutA}[1]{}

\usepackage{graphicx}
\begin{document}

\title{Graph-based linear scaling electronic structure theory}

\author{Anders M.\ N.\ Niklasson}
\affiliation{Theoretical Division, Los Alamos National Laboratory, Los Alamos, New Mexico 87545}
\author{Susan M.\ Mniszewski}
\affiliation{Computer, Computational, and Statistical Sciences Division, Los Alamos National Laboratory, Los Alamos, New Mexico 8754}
\author{Christian F.\ A.\ Negre}
\affiliation{Theoretical Division, Los Alamos National Laboratory, Los Alamos, New Mexico 87545}
\author{Marc J.\ Cawkwell}
\affiliation{Theoretical Division, Los Alamos National Laboratory, Los Alamos, New Mexico 87545}
\author{Pieter J.\ Swart}
\affiliation{Theoretical Division, Los Alamos National Laboratory, Los Alamos, New Mexico 87545}
\author{Jamal Mohd-Yusof}
\affiliation{Computer, Computational, and Statistical Sciences Division, Los Alamos National Laboratory, Los Alamos, New Mexico 8754}
\author{Timothy C.\ Germann}
\affiliation{Theoretical Division, Los Alamos National Laboratory, Los Alamos, New Mexico 87545}
\author{Michael E.\ Wall}
\affiliation{Computer, Computational, and Statistical Sciences Division, Los Alamos National Laboratory, Los Alamos, New Mexico 8754}
\author{Nicolas Bock}
\affiliation{Theoretical Division, Los Alamos National Laboratory, Los Alamos, New Mexico 87545}
\author{Emanuel H. Rubensson}
\affiliation{Division of Scientific Computing, Department of Information Technology, Uppsala University, Box 337, SE-751 05 Uppsala, Sweden}
\author{Hristo Djidjev}
\affiliation{Computer, Computational, and Statistical Sciences Division, Los Alamos National Laboratory, Los Alamos, New Mexico 87545}

\date{\today}

\begin{abstract}
{We show how graph theory can be combined with quantum theory to calculate the electronic
structure of large complex systems. The graph formalism is general
and applicable to a broad range of electronic structure methods
and materials, including challenging systems such as biomolecules.
The methodology combines well-controlled accuracy, low computational cost,
and natural low-communication parallelism. This combination addresses
substantial shortcomings of linear scaling electronic structure theory,
in particular with respect to quantum-based molecular dynamics simulations.}
\end{abstract}


\maketitle

\section{Introduction}

The importance of electronic structure theory in materials science,
chemistry, and molecular biology, relies on the development of
theoretical methods that provide sufficient accuracy at a reasonable computational cost.
Currently, the field is dominated by Kohn-Sham density functional theory
\cite{hohen,WKohn65,RParr89,RMDreizler90}, which often combines good theoretical fidelity
with a modest computational workload that is constrained mainly by the diagonalization of
the Kohn-Sham Hamiltonian -- an operation that scales cubically with
the system size. However, for systems beyond a few hundred
atoms, the diagonalization becomes prohibitively expensive.
This bottleneck was removed with the development
of linear scaling electronic structure theory \cite{SGoedecker99,DBowler12},
which allows calculations of systems with millions of atoms \cite{DBowler10,JVandevondele12}.
Unfortunately, the immense promise of linear scaling electronic structure
theory has never been fully realized because of some significant shortcomings, in particular,
$a)$ the accuracy is reduced to a level that is often
difficult, if not impossible, to control;
$b)$ the computational pre-factor is high and
the linear scaling benefit occurs only for very large systems that
in practice often are beyond acceptable time limits or available computer resources; and
$c)$ the parallel performance is generally challenged by a significant
overhead and the wall-clock time remains high even with massive parallelism.
In quantum-based molecular dynamics (QMD) simulations \cite{DMarx00} 
all these problems coalesce and we are constrained either 
to small system sizes or short simulation times.

In this paper we propose to overcome these shortcomings by introducing a 
formalism based on graph theory \cite{GChartrand85,JABondy08} that allows
practical and easily parallelizable electronic structure calculations
of large complex systems with well-controlled accuracy.
The graph-based electronic structure theory
combines the natural parallelism of a divide and conquer approach
\cite{WYang91,PDWalker93,WYang95,IAbrikosov96,KKitaura99,TOzaki06}
with the automatically adaptive and tunable accuracy of a thresholded sparse matrix algebra
\cite{FGustavson78,SPissanetzky84,WPress92,YSaad96,MChallacombe00B,EHRubensson07,EHRubensson08,ABuluc12,UBorstnik14,NBock14,VWeber15,SMniszewski15,MKL,cuSPARSE},
which can be combined with fast, low pre-factor, recursive Fermi operator expansion methods
\cite{RMcWeeny56,APalser98,AHolas01,ANiklasson02,ANiklasson03B,WLiang03,ERudberg11,EHRubensson11,PSuryanarayana13,EHRubensson14}
and be applied to modern formulations of Born-Oppenheimer molecular dynamics (MD)
\cite{PPulay04,JMHerbert05,ANiklasson06,TDKuhne07,GZheng11,JHutter12,LLin14,MArita14,ANiklasson14}.

\section{Results}

\subsection{Graph-based electronic structure theory}

\begin{figure*}[ht]
\begin{center}
\centerline{\includegraphics[width=.6\textwidth]{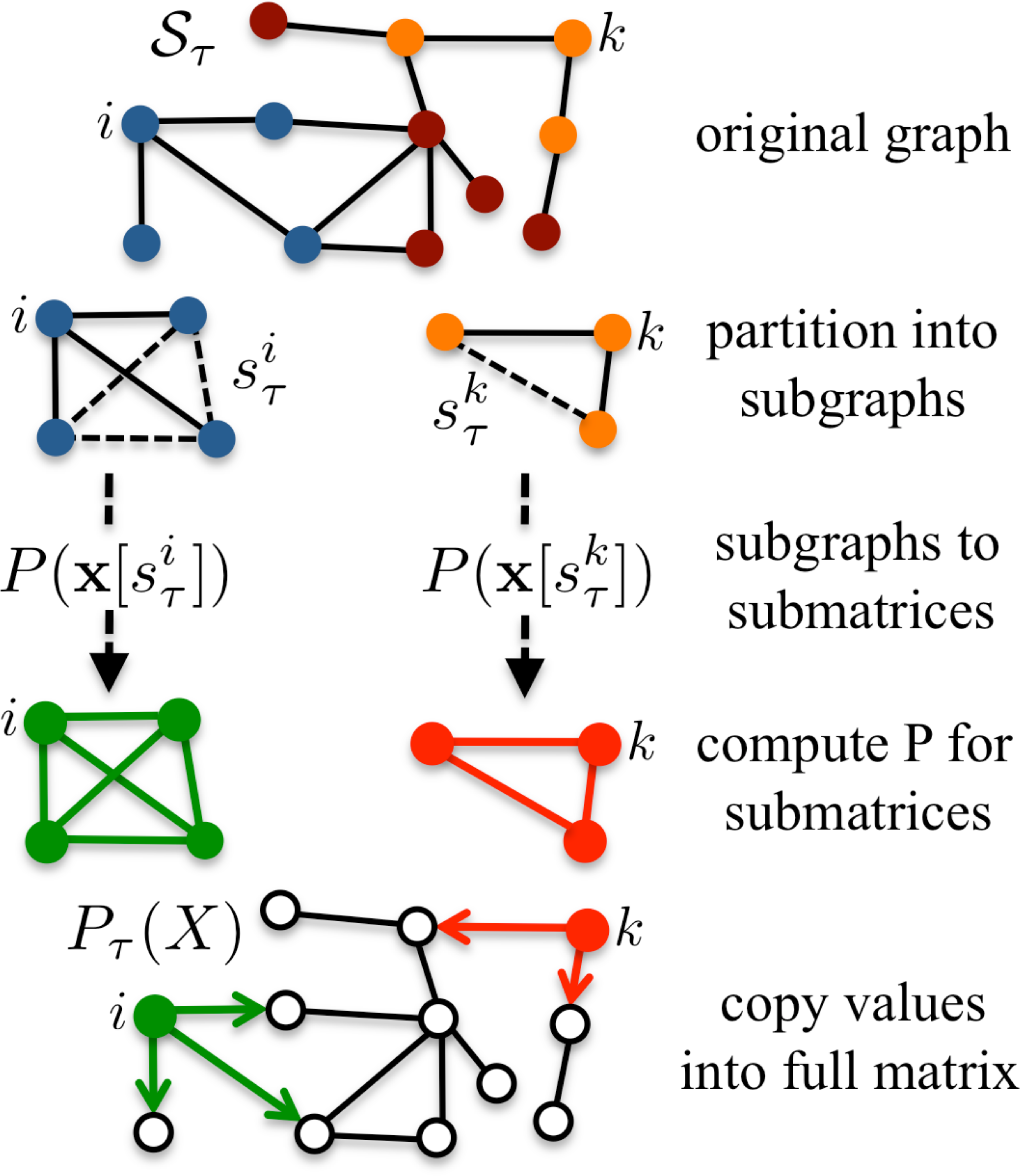}}
\caption{The data dependency graph ${\cal S}_\tau$
and the subgraphs ($s_\tau^i$ or $s_\tau^k$),
one for each {\em core} vertex ($i$ or $k$) including
all directly connected {\em halo} vertices in ${\cal S}_\tau$.
The full matrix polynomial $P_\tau(X)$ is given by an assembly from
$P({\bf x}[s_{\tau}^i])$ of the separate dense subgraph
contractions ${\bf x}[s_\tau^i]$.}\label{Fig1}
\end{center}
\end{figure*}

Our graph-based electronic structure theory relies on the equivalence between the
calculation of thresholded sparse matrix polynomials and a graph partitioning approach.
Let $P(X)$ be a $M$th-order polynomial of a $N\times N$ symmetric square matrix $X$ that
is given as a linear combination of some basis polynomials $T^{(n)}(X)$,
\begin{equation}\label{Pol}\begin{array}{l}
{\displaystyle P(X) = \sum_{n = 0}^{M} c_nT^{(n)}(X)}.
\end{array}
\end{equation}
We define an approximation $P_\tau (X)$ of $P(X)$ using a {\em globally}
thresholded sparse matrix algebra,
where matrix elements with a magnitude below a numerical threshold
$\tau$ {\em in all terms}, $T^{(n)}(X)$, are ignored.
The pattern of the remaining matrix entries, which at any point of the expansion have been
(or are expected to be) greater than $\tau$, can be described by a \textit{data dependency graph}
${\cal S}_\tau$ that represents all possible data dependencies between the matrix elements
in the polynomial expansion.
Formally, we define the graph $S_\tau$ with a vertex for each row of $X$ and
an edge $(i,j)$ between vertices $i$ and $j$ if
\begin{equation} \label{DataConn}
\{T^{(n)}(X)\}_{i,j}\geq \tau \mbox{ for any $n\leq M$.}
\end{equation}
For a matrix $A$, we denote by $\lfloor A \rfloor_{{\cal S}_\tau}$
the thresholded version of $A$, where
\begin{equation} \label{thresh_mat}
\left\{\lfloor A \rfloor_{{\cal S}_\tau}\right\}_{i,j} = \left\{ \begin{array} {ll}
A_{i,j} & \mbox{ if $(i,j)$ is an edge of $S_\tau$} \\
0 & {\rm otherwise.} \end{array}
\right.
\end{equation}
The thresholded polynomial $P_{\tau}(X)$ of $P(X)$ with respect to ${\cal S}_\tau$ is given by
\begin{equation}\label{Ptau}
P_{\tau}(X)= \sum_{n = 0}^{M} c_nT^{(n)}_{{\cal S}_\tau}(X),
\end{equation}
where the thresholded $T^{(n)}_{{\cal S}_\tau}(X)$ can be calculated from a linear recurrence
\begin{equation}\label{Truncation}
{\displaystyle T^{(n)}_{{\cal S}_\tau}(X) =
\alpha_n\lfloor X T^{(n-1)}_{{\cal S}_\tau}(X)\rfloor_{{\cal S}_\tau}
+ \sum_{m=0}^{n-1} \alpha_m T^{(m)}_{{\cal S}_\tau}(X)},
\end{equation}
with $T^{(0)}_{{\cal S}_\tau}(X) = I$.
A key observation of this paper is that the calculation of $P_{\tau}(X)$
in Eqs.\ (\ref{Ptau}) and (\ref{Truncation})
is equivalent to a partitioned subgraph expansion on ${\cal S}_\tau$.
This approach is illustrated in Fig.\ \ref{Fig1}.
For any vertex $i$ of ${\cal S}_\tau$, let $s_\tau^i$ be the
subgraph of ${\cal S}_\tau$
induced by the {\em core} (meaning belonging to a single subgraph)
vertex $i$ and all {\em halo} (shared) vertices that
are directly connected to $i$ in ${\cal S}_\tau$.
Then the $i$th matrix column of $P_{\tau}(X)$
is given by the thresholded expansion determined by $s_\tau^i$ only, i.e.
\begin{equation}\label{GraphPart}
{\displaystyle \left\{P_{\tau}(X)\right\}_{:,i} = \left\{P({\bf x}[s_{\tau}^i])\right\}_{:,j}~}.
\end{equation}
Here $j$ is the column (or row) of the polynomial for the subgraph
$s_\tau^i$ containing all edges from the core vertex $i$ that corresponds to column $i$
of the complete matrix polynomial on the left-hand side. ${\bf x}[s_{\tau}^i]$ is the
small dense principal submatrix that contains only the entries of $X$ corresponding to $s_{\tau}^i$.
The full matrix  $P_{\tau}(X)$ can then be assembled, column by column,
from the set of smaller dense matrix polynomials $P({\bf x}[s_{\tau}^i])$
for each vertex $i$. The calculation of a numerically thresholded matrix polynomial $P_{\tau}(X)$
thus can be replaced by a sequence of fully independent small
dense matrix polynomial expansions determined by a graph partitioning.

Equation (\ref{GraphPart}) represents an exact relation between
a globally thresholded sparse matrix algebra and a graph partitioning approach,
which is valid for a general matrix polynomial $P(X)$.
Several observations can be made about this equivalence:
$i)$ $P_{\tau}(X)$ is not symmetric and with the order of the matrix product
for the threshold in Eq.\ (\ref{Truncation}) we collect $P_{\tau}(X)$
column by column in Eq.\ (\ref{GraphPart}) as illustrated by the {\em directed} graph at the bottom of Fig.\ 1;
$ii)$ the accuracy of the matrix polynomial
increases (decreases) as the threshold $\tau$ is reduced (increased)
and the number of edges of ${\cal S}_\tau$ increases (decreases);
$iii)$ we may thus include additional edges in ${\cal S}_\tau$ without loss of accuracy;
$iv)$ the polynomial $P_{\tau}(X)$ is zero at all entries outside of ${\cal S}_\tau$;
$v)$ apart from spurious cancellations, the non-zero pattern of $P_{\tau}(X)$ is therefore the same as ${\cal S}_\tau$
and we can expect a numerically thresholded exact matrix polynomial, $\lfloor P(X)\rfloor_\tau$, to
have a non-zero structure similar to ${\cal S}_\tau$;
$vi)$ the graph partitioning can be generalized such that each vertex
corresponds to a combined set of vertices, i.e.\ a community, without loss of accuracy;
$vii)$ we may reduce the computational cost by identifying such communities using graph partitioning schemes,
which can be tailored for optimal platform-dependent performance;
$viii)$ the exact relation given by Eqs.\ (\ref{Ptau}) - (\ref{GraphPart}) holds for any structure
of ${\cal S}_\tau$ and is not limited to the threshold in Eq.\ (\ref{DataConn});
$ix)$ the particular sequence of matrix operations in the calculation
of $P_{\tau}(X)$ is of importance because of the thresholding in Eq.\ (\ref{Truncation}),
whereas the order (or grouping) of the matrix multiplications is arbitrary for the contracted
matrix polynomials $P({\bf x}[s_{\tau}^i])$ in Eq.\ (\ref{GraphPart}); and
$x)$ the computational cost of each polynomial expansion
is dominated by separate sequences of dense matrix-matrix multiplication that can be performed
independently and in parallel.

A main point of this paper is that the equivalence between the calculation of the thresholded
sparse matrix polynomial and the graph partitioned expansion in Eq.\ (\ref{GraphPart}) provides
a natural framework for a graph-based formulation of linear scaling electronic structure theory.
In Kohn-Sham density functional theory the matrix polynomial in Eq.\ (\ref{Pol})
is replaced by the Fermi-operator expansion \cite{RParr89,SGoedecker94,RSilver94} where
\begin{equation}\label{FOE}
P(H) = D = \left[ e^{\beta(H-\mu)} + 1\right]^{-1} \approx \sum_{n=0}^M c_n T^{(n)}(H).
\end{equation}
Here $D$ is the density matrix, $H$ the Hamiltonian,
$\mu$ the chemical potential, and $\beta$ the inverse temperature.
The matrix functions, $T^{(n)}(X)$,
are typically Chebyshev polynomials constructed by a
recurrence equation as in Eq.\ (\ref{Truncation}).  With a local basis set
$H$ and $P(H)$ have
sparse matrix representations above some numerical threshold
for sufficiently large non-metallic systems \cite{SGoedecker99,DBowler12}.
The graph-based construction of sparse matrix polynomials in Eq.\ (\ref{GraphPart})
can then be applied to the calculation of the density matrix with the data dependency
graph ${\cal S}_\tau$ estimated from an approximate prior density matrix that is available
in an iterative self-consistent field (SCF) optimization
or from previous time steps in a MD simulation.
The computation can be accelerated with a recursive Fermi-operator expansion
\cite{RMcWeeny56,APalser98,AHolas01,ANiklasson02,ANiklasson03B,WLiang03,EHRubensson11,PSuryanarayana13,EHRubensson14}.
In the zero temperature limit the Fermi function equals the Heaviside step function $\theta$ and
a recursive expansion is then given by
$D = \theta(\mu I-H) = \lim_{n \rightarrow \infty} f_n(f_{n-1}(\ldots f_0(H)\ldots ))$, which
reaches a high expansion order much more rapidly compared to the serial form in Eq.\ (\ref{Pol}).
With $f_n(X)$ being 2nd-order polynomials \cite{ANiklasson02} we reach an expansion order of over a billion in only 30 iterations.
The ability to use a fast recursive expansion is motivated from $ix)$ above, and since
any recursive expansion also can be written in the general form of Eq.\ (\ref{Pol}).
Once the density matrix $D$ is known, the expectation value of any operator $A$ is given
by $\langle A \rangle = Tr[DA]$. Generalizations to quantum perturbation theory are straightforward 
\cite{ANiklasson04,VWeber04}.

\subsection{Macromolecular test system}

\begin{figure*}[ht]
\begin{center}
\centerline{\includegraphics[width=.7\textwidth]{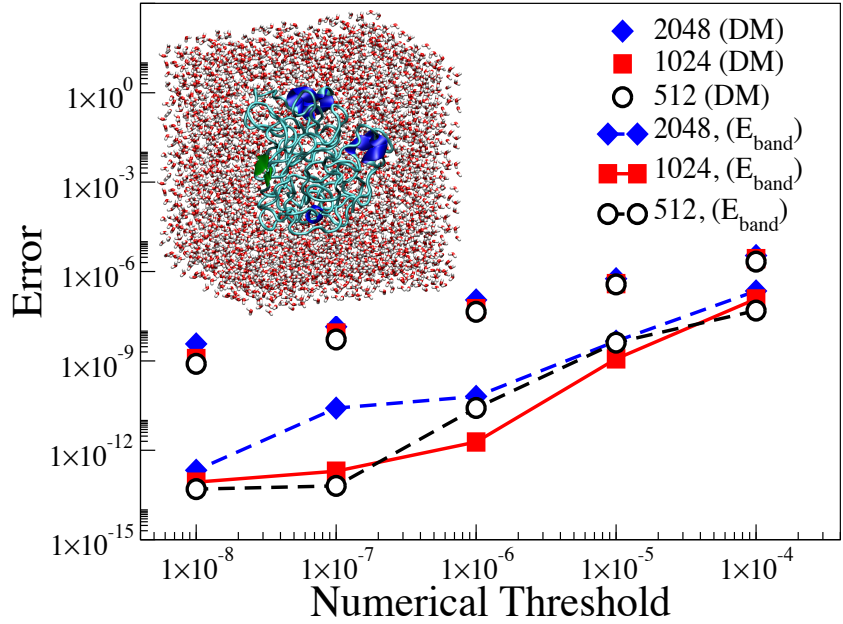}}
\caption{The error in the calculated density matrix (DM)
for polyalanine (2,593 atoms) in water with a total of 19,945 atoms (inset),
as measured by the Frobenius norm (normalized per atom)
for partitions with 512, 1024, and 2048 separate communities based on
graphs, ${\cal S}_\tau$, from varying numerical thresholds $\tau$.
The connected symbols (lower part) shows the error in 
band energy, $E_{\rm band} = Tr[HD]$, per atom.}\label{Fig2}
\end{center}
\end{figure*}

\begin{figure*}[ht]
\begin{center}
\centerline{\includegraphics[width=.7\textwidth]{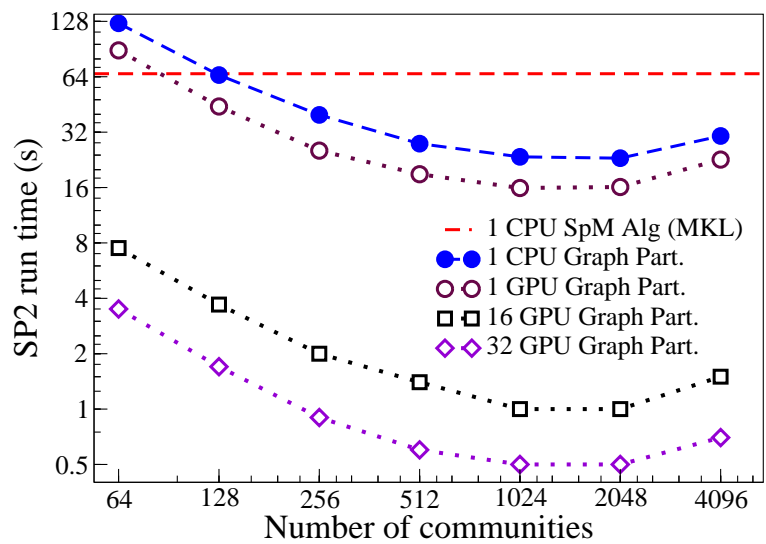}}
\caption{The time to calculate the density matrix using the SP2
expansion (with threshold $\tau = 10^{-5}$) partitioned over different sets of subgraphs for the
solvated polyalanine system (19,945 atoms).}\label{Fig3}
\end{center}
\end{figure*}

Figure \ref{Fig2} shows the error per atom in the density matrix of the band energy, $E_{\rm band} = Tr[DH]$,
calculated with the graph-based formulation above for a 19,945-atom
macromolecular system of polyalanine solvated in water.
The calculations were performed using self-consistent charge density functional
tight-binding theory \cite{MElstner98,MFinnis98,TFrauenheim00}
as implemented in the electronic structure program LATTE \cite{MCawkwell12} in combination with
the recursive second-order spectral projection (SP2) zero-temperature Fermi-operator expansion scheme \cite{ANiklasson02}.
The data dependency graphs, ${\cal S}_\tau$, were estimated
by thresholding an ``exact'' density matrix with varying thresholds, $\tau$.
Different numbers of subgraph communities (512, 1024 or 2048)
were chosen and optimized with the METIS heuristic multilevel graph partitioning package \cite{metis}
for the different data dependency graphs (one for each threshold)
using the multilevel recursive bisection method. The errors
were determined in comparison to
the ``exact'' density matrix, which was calculated using regular sparse matrix
algebra with a tight threshold of $10^{-12}$.
The error is fairly insensitive to the number of graph partitions and is instead controlled
by the value of the threshold that is used to estimate the data dependency graphs.
In contrast, the computational cost varies significantly with the size of the graph partitions.
The cost in the limit of only one large community, containing the whole system,
or in the opposite limit, with one partition for each orbital,
scale as ${\cal O}(N^3)$ or ${\cal O}(Nm^3)$, respectively,
where $m$ is the average number of edges per vertex in ${\cal S}_\tau$
and $N \times N$ is the size of $H$.
A straightforward graph partitioning may thus lead to a significant
overhead compared to a Fermi-operator expansion using thresholded sparse
matrix algebra \cite{SGoedecker99}, which scales as ${\cal O}(Nm^2)$.
However, with an optimized graph partitioning the total cost can
be reduced to scale as ${\cal O}(Nm^2)$.
A similar optimization can be performed for divide and conquer methods,
but may not be applicable to inhomogeneous systems \cite{TOzaki06}.
Figure~\ref{Fig3} shows the timing (66 s, red dashed line)
for a thresholded sparse matrix algebra (SpM Alg) Fermi-operator expansion
with Intel's MKL sparse matrix library \cite{MKL} running in parallel
on a dual eight-core CPU.
With the graph-based approach (filled circles) using
the METIS graph partitioning (Graph Part.)
program for varying numbers of communities
it is possible to significantly reduce the run time on the same platform (23 s).
The graph-based formalism also has the additional advantage of
an almost trivial and highly scalable parallelism as is demonstrated
by the run times on 1, 16 or 32 graphics processing units (GPUs)
on separate nodes (open symbols) \cite{cuBLAS}.
The parallel performance is close to ideal, reaching a performance
of about 25 $\mu s$/atom and a subsecond wall-clock time on the 32 node GPU platform.


\subsection{Molecular dynamics simulation}
Linear scaling divide and conquer methods \cite{WYang91,PDWalker93,WYang95,IAbrikosov96,KKitaura99,TOzaki06}
rely on an estimated finite range of direct electron interaction, which can be motivated by
the localized character of the Wannier functions \cite{WKohn96,WKohn64,Mott61}.
This allows a system to be partitioned into smaller overlapping regions
that are solved separately (apart from long-range electrostatic interactions),
within pre-determined local interaction zones, and then reassembled.
Divide and conquer schemes are naturally parallel and in spirit
similar to our graph-based approach. However, their numerical accuracy
can be difficult to control without careful prior testing and convergence
analysis \cite{TSLee96,DMYork98,DBowler12}.  An automatic, adjustable error control is
particularly challenging in MD simulations of inhomogeneous materials,
where reacting or floppy molecules and atoms can move across pre-determined
local interaction zones and where transitions between localized and
itinerant electronic states may occur.
MD simulations of inhomogeneous molecular systems with significant changes in the electronic overlap are therefore
of particular interest when we evaluate our framework. Furthermore,
the precision can be gauged very sensitively by the accuracy and long-term stability of the total energy.

The data dependency graph ${\cal S}_\tau(t)$ can be estimated from the numerically thresholded density matrix
in the previous MD time step, $\lfloor D(t-\delta t)\rfloor_\tau$,
and new Hamiltonian matrix elements, $H(t)$, as the atoms move, for example, from
\begin{equation}\label{S_est}
{\cal S}_\tau(t) \leftarrow \lfloor \left( \lfloor D(t-\delta t) \rfloor_\tau + H(t)\right) ^2 \rfloor_\epsilon.
\end{equation}
In our MD simulation below we use the symbolic representation
of ${\cal S}_\tau(t)$ in Eq.\ (\ref{S_est}), which is given from the non-zero pattern of
the thresholded density matrix (with $\tau = 10^{-4}$) combined
with the non-zero pattern of $H(t)$, and instead of the matrix square we use
paths of length two, corresponding to the symbolic operation ($\epsilon = 0$).
This approach that adapts ${\cal S}_\tau(t)$
to each new MD time step by including additional redundant 
edges works surprisingly well, though ${\cal S}_\tau(t)$ cannot 
increase by more than paths of length two between two MD steps.
Generalized estimates can also be designed for the iterative SCF optimization.

\begin{figure*}[ht]
\begin{center}
\centerline{\includegraphics[width=.7\textwidth]{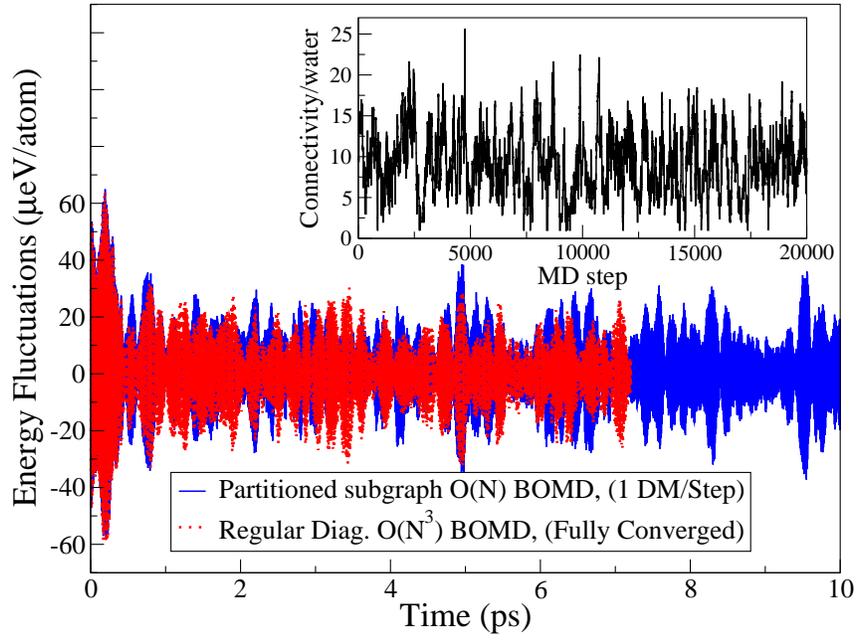}}
\caption{The total energy fluctuations in a microcanonical Born-Oppenheimer MD (BOMD)
simulation of liquid water (100 molecules, $T \sim 300$ K, $\delta t = 0.5$ fs),
using graph partitioning and one density matrix (DM) construction per step vs.
SCF optimized BOMD with diagonalization (Diag.).
The inset shows the number of water molecules associated with the subgraph
of an individual molecule. Energy drift is less than $\sim 0.2 \mu$eV/atom per ps.}\label{Fig4}
\end{center}
\end{figure*}

Figure \ref{Fig4} shows the fluctuations of the total energy
during a microcanonical MD simulation of liquid water that
was performed using LATTE \cite{MCawkwell12}
and the extended Lagrangian formulation of Born-Oppenheimer MD
\cite{ANiklasson08,PSteneteg10,PSouvatzis14,ANiklasson14,BAradi15}.
The density matrix was calculated from a partitioning over separate subgraphs of ${\cal S}_\tau(t)$, with one
water molecule per core.
For the Fermi-operator expansion (at zero temperature) we used the recursive SP2 algorithm \cite{ANiklasson02}.
In each time step the complete SP2 sequence
(the same for each subgraph expansion) for the correct total occupation is pre-determined
from the HOMO-LUMO gap that is estimated from the previous time step as in Ref.~\cite{EHRubensson14}.
In this way each expansion can be performed independently, without exchange of
information during or between the matrix multiplications as otherwise would be required
\cite{JVandevondele12,VWeber15}. Communication is
reduced to a minimum and no additional adjustments of the electronic occupation,
as in divide and conquer calculations \cite{WYang95}, is required.
The inset of Fig.\ (\ref{Fig4}) shows the number of water molecules of a single subgraph (core + halo)
along the trajectory of an individual molecule,
which oscillates as ${\cal S}_\tau(t)$
adaptively follows the fluctuations in the electronic overlap.
Despite the large oscillations, including
between 1 and 25 molecules, the total energy is both accurate and stable. The ``exact''
calculation with fully converged density matrices ($\ge$4 SCFs per step) using
dense matrix algebra based on full ${\cal O}(N^3)$ diagonalization,
is virtually indistinguishable for the first 0.5 ps (or 1,000 time steps).

\section{Discussion}

\begin{figure*}[ht]
\begin{center}
\centerline{\includegraphics[width=.7\textwidth]{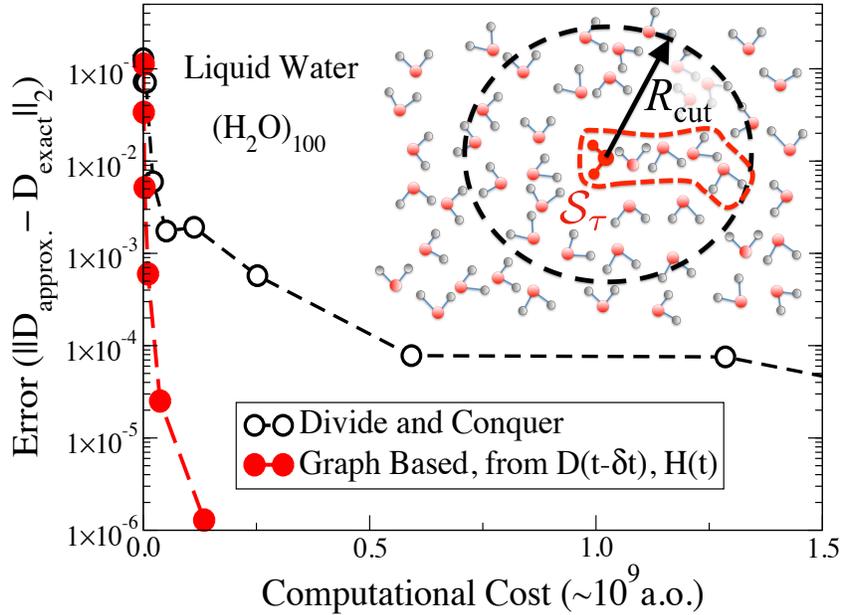}}
\caption{The convergence of the density matrix error for a snapshot during a MD simulation
of the water system in Fig.\ \ref{Fig4} ($100$ molecules, $T \sim 300$ K, $\delta t = 0.5$ fs)
as a function of the computational cost for various numerical
thresholds ($\tau = 10^{-1},10^{-2}, \ldots, 10^{-6}$)
in the symbolic estimate of the data-dependency graph in Eq.\ (\ref{S_est})
for the graph-based method, and for different sizes of the cutoff radius, $R_{\rm cut}$, 
in a divide and conquer approach. To capture a hypothetical electronic overlap 
within the red dashed border in the inset (associated with the data-dependency graph ${\cal S}_\tau$
for the large red molecule at the center),
the cutoff radius needs to be large, which leads to a significant overhead for the divide and conquer approach. 
The efficiency would be similar only for a homogeneous system. 
The computational cost was estimated from the sum 
of the number arithmetic operations (a.o.) required to calculate the density matrices 
from all the separate subgraph partitions or divide and conquer regions -- 
one for each water molecule.}\label{Fig5}
\end{center}
\end{figure*}

Linear scaling QMD simulations using divide and conquer or radial truncation approaches
often show systematic energy drifts \cite{FShimojo08,ETsuchida08,FShimojo14}
that are significantly higher than regular ${\cal O}(N^3)$ methods \cite{DMarx00,PPulay04,JMHerbert05}
and multiple orders of magnitude larger than the graph-based QMD simulation in Fig.\ \ref{Fig4}.
Such problems may occur because of difficulties controlling
the error in the force evaluations \cite{DBowler12,MKobayashi11} as atoms
move across the local zone boundaries and as the electronic overlap fluctuates,
or because of incomplete SCF optimization causing a broken time-reversal symmetry \cite{DRemler90,PPulay04}.
The problem is illustrated in Fig.\ \ref{Fig5}, which shows a comparison between a divide and conquer
approach and our graph-based calculation of the density matrix for a snapshot from a MD simulation of the water system in Fig.\ \ref{Fig4}.
Without the adaptivity of the graph-based method, the divide and conquer approach needs a large cutoff
radius, $R_{\rm cut}$, to reach sufficient convergence in the calculation of the density matrix for the water system, 
which leads to a significant overhead. With the graph-based framework as demonstrated here
in combination with a modern formulation of Born-Oppenheimer MD
\cite{PPulay04,JMHerbert05,ANiklasson06,TDKuhne07,GZheng11,JHutter12,LLin14,ANiklasson14,MArita14},
these problems can be avoided.

The off-the-shelf METIS graph partitioning scheme used for the macromolecular system
in Figs.\ \ref{Fig2} and \ref{Fig3} works very
well and drastically reduces the overhead compared to a straightforward implementation.
However, by adjusting the graph partitioning to the particular requirements of
the electronic structure calculation as well as the computational platform,
further optimizations are possible \cite{Djidjev}. We may also incorporate
an on-the-fly graph partitioning that updates the subgraphs during a MD 
simulation as the structure changes.

\section{Conclusions}

In this article we have shown how graph theory can be combined with quantum theory to calculate the electronic
structure of large complex systems with well-controlled accuracy.  The graph formalism is general and
applicable to a broad range of electronic structure methods and materials, for which sparse matrix representations
can be used, including QMD simulations, overcoming significant gaps in
linear scaling electronic structure theory.

\begin{acknowledgments}
We acknowledge support from Office of Basic Energy Sciences (LANL2014E8AN)
and the Laboratory Directed Research and
Development program of Los Alamos National Laboratory (LANL).
Generous support and discussions with T. Peery at the
T-division International Java Group are acknowledged.
The research used resources provided by the LANL
Institutional Computing Program. LANL, an
affirmative action/equal opportunity employer, is operated by
Los Alamos National Security, LLC, for the National Nuclear
Security Administration of the U.S. DOE under Contract No.
DE-AC52-06NA25396.
\end{acknowledgments}

\bibliography{mondo_new_x}


\newpage

\newpage

\newpage

\newpage


\newpage

\end{document}